\documentclass[9pt,a4paper,twocolumn]{article}
\input{pisika.dat}
\begin{document}

\providecommand{\ShortAuthorList}[0]{}
\title{Universal Unitary Photonic Circuits by
Interlacing Discrete Fractional Fourier Transform and
Phase Modulation}
\author[1,2]{Matthew Markowitz} 
\author[1,2,*]{Mohammad-Ali Miri}
\affil[1]{Department of Physics, Queens College of the City University of New York, Queens, New York 11367, USA}
\affil[2]{Physics Program, The Graduate Center, City University of New York, New York, New York 10016, USA}
\affil[*]{Corresponding author: mmiri@qc.cuny.edu}

\begin{abstract}
\noindent

We introduce a novel parameterization of complex unitary matrices, which allows for the efficient photonic implementation of arbitrary linear discrete unitary operators. The proposed architecture is built on factorizing an $N \times N$ unitary matrix into interlaced discrete fractional Fourier transforms and $N$-parameter diagonal phase shifts. We show that such a configuration can represent arbitrary unitary operators with $N+1$ phase layers. We discuss a gradient-based algorithm for finding the optimal phase parameters for implementing a given unitary matrix. By increasing the number of phase layers beyond the critical value of $N+1$, the optimization consistently converges faster as the system becomes over-determined. We propose an integrated photonic circuit realization of this architecture with coupled waveguide arrays and reconfigurable phase modulators. The proposed architecture can pave the way for developing novel families of programmable photonic circuits for optical classical and quantum information processing.

\DOI{}
\end{abstract}

\maketitle
\thispagestyle{titlestyle}
A universal photonic device that can perform arbitrary discrete linear unitary operations is an indispensable component for numerous applications in classical and quantum optics \cite{bogaerts2020programmable}. Such a component provides a great platform for multistate systems of quantum particles for quantum computation and quantum information processing \cite{schaeff2015experimental, carolan2015universal,  harris2017quantum}. In classical optical computing, a programmable matrix-vector multiplier is critical for realizing photonic artificial neural networks among other unconventional computing paradigms \cite{shen2017deep,harris2018linear}. On the other hand, an integrated photonic matrix-vector multiplier is intrinsically a universal multiport linear device with a wide range of applications in analog signal processing, communications, and sensing \cite{miller2013self, perez2017multipurpose}.

In essence, the physical realization of a programmable device that performs arbitrary unitary operations requires a suitable parameterization and factorization of unitary matrices into matrix factors with fewer parameters \cite{dita2003factorization} that can be effectively implemented with photonic components. In a pioneering work such factorization has been recursively utilized for representing unitary operators in a feedforward triangular mesh architecture of beam splitters and phase shifters \cite{reck1994experimental}. This architecture can be implemented with integrated photonic components, resulting in programmable circuits which for an $N \times N$ unitary matrix require $N^2$ phase shifters and $N(N-1)/2$ Mach–Zehnder interferometers \cite{miller2013self, crespi2013integrated, ribeiro2016demonstration, harris2016large, taballione20198}. It is shown that an alternative arrangement of the underlying beam splitters and phase shifters through a rectangular architecture reduces the overall length of the design \cite{clements2016optimal}. On the other hand, it is shown that $N \times N$ unitary operations can be universally realized through a fundamentally different architecture that is built on cascading $N$-point discrete Fourier transforms (DFT) with arrays of $N$ phase shifters \cite{morizur2010programmable, armstrong2012programmable, pastor2021arbitrary}. In particular, a deterministic algorithm is provided to decompose a unitary matrix with $6N$ DFT layers and $6N+1$ phase layers. However, this architecture results in large device lengths primarily because of the large number of cascaded layers while on-chip implementation of DFT itself is difficult and requires bulky multimode interference (MMI) devices \cite{pastor2021arbitrary}. Here, we propose a universal cascaded architecture that requires only $N+1$ phase layers and guarantees one order-of-magnitude size reduction in an integrated photonics implementation.

\begin{figure*}[ht]
\centering 
\includegraphics[width=0.9\textwidth]{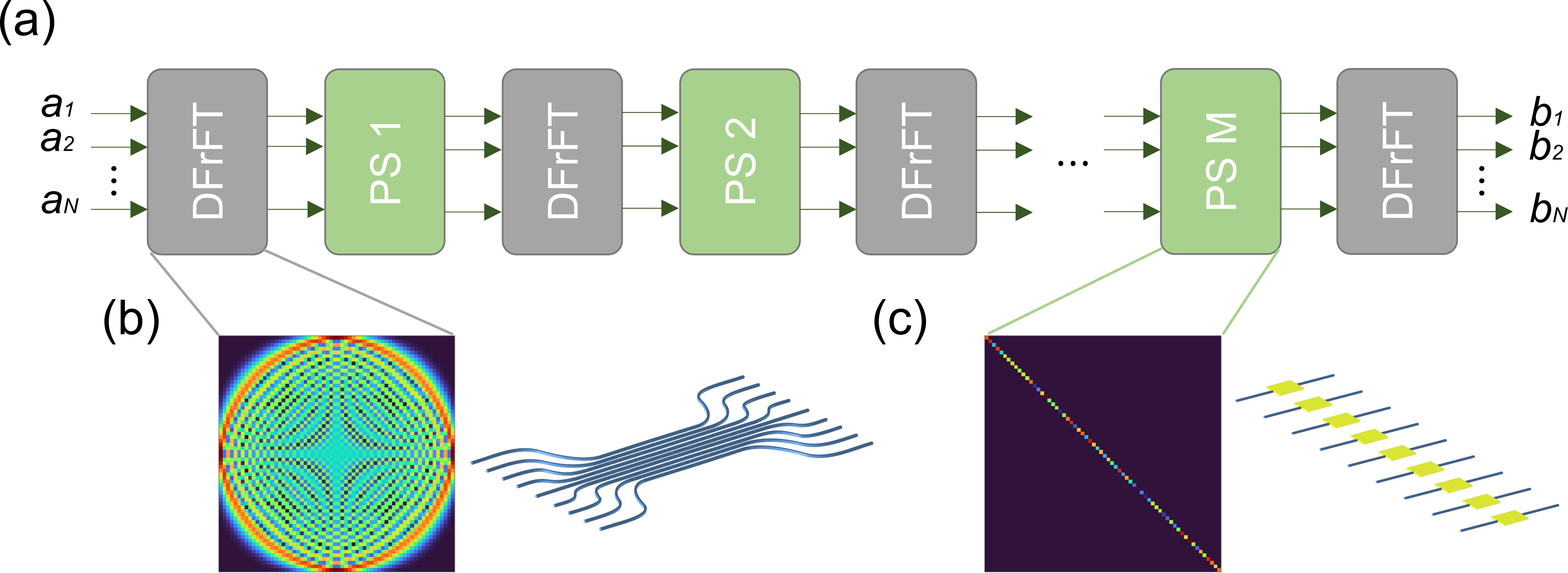}
\caption{(a) The proposed architecture of the N-port system, consisting of alternating layers of Discrete Fractional Fourier Transforms (DFrFT) and diagonal phase shifts (PS). (b,c) Schematics showing matrix representations of DFrFT and PS operations and their physical realizations as integrated photonic components.}
\label{fig1}
\end{figure*}

In this letter, we introduce a novel architecture for realizing unitary discrete linear operators. The proposed architecture is built on interlacing phase shifter arrays with Discrete Fractional Fourier Transform (DFrFT) operators as shown in Fig.~\ref{fig1}. The fractional Fourier transform can be realized with photonic waveguide arrays \cite{weimann2016implementation,HonariLatifpour2022}, which allows for the integrated photonic realization of such an architecture by using two basic elements, i.e., waveguides and phase shifters as illustrated in Fig.~\ref{fig1}. Through numerical investigations, we show that this architecture universally represents unitary matrices with only $N+1$ phase layers. To investigate the universality of this architecture we consider ensembles of $N \times N$ random unitary matrices and explore their approximation with the proposed architecture by considering a finite number of phase layers and finding the optimal set of phase parameters through rigorous gradient-based optimization. In each case, we explore the norm of error for each matrix and its representation with the proposed architecture versus the number of phase layers. Our results indicate an abrupt phase transition in the norm of error at exactly $N$ phase layers.

Considering a general $N \times N$ unitary transformation matrix $U$ ($U^{\dagger} U = U U^{\dagger} = I$), our goal is to find a representation in the following form (as visualized in Fig.~\ref{fig1}):
\begin{equation}
\label{eq_U}
    U = F P_{M} F \cdots P_m \cdots F P_1 F
\end{equation}
where, $P_m$ ($m=1,2,3,\cdots,M$) are diagonal phase matrices, and $F$ is the Discrete Fractional Fourier Transform (DFrFT). The phase matrix is defined as
\begin{equation}
P_m = \begin{bmatrix}
      e^{i \theta_1^m} & 0 & \dots & 0\\
      0 & e^{i \theta_2^m} & \ddots & \vdots\\
      \vdots & \ddots & \ddots & 0\\
      0 & \cdots & 0 & e^{i \theta_N^m}
      \end{bmatrix}
\qquad
\end{equation}

where, $m=1,\cdots,M$, refers to the phase layer, and $\theta_i^m$ represents the $i$th phase element in the $m$th layer of phase shifter arrays. It is important to note that there are several valid definitions of the discrete fractional Fourier transform. Here, we adapt the definition from Ref.~\cite{weimann2016implementation} which allows a physical realization of the DFrFT with a particular photonic waveguide array, called the Jx array. In this fashion, the DFrFT matrix of fractional order $\alpha=\pi/2$ can be written as the propagator operator of such a lattice for the normalized length $\pi/2$:

\begin{equation}
\label{eq_F}
F = e^{i\frac{\pi}{2}H},
\end{equation}
where, $H$ is the Hamiltonian of the Jx lattice
\begin{equation}
\label{eq_H}
H = \begin{bmatrix}
    0 & \kappa_{1,2} & \dots & 0\\
     \kappa_{1,2} & 0 & \ddots & \vdots \\
    \vdots & \ddots & \ddots & \kappa_{N-1,N}\\
    0 & \dots & \kappa_{N-1,N} & 0
    \end{bmatrix}
\qquad
\end{equation}
while the hopping rates are given by \cite{weimann2016implementation}
\begin{equation}
\label{eq_kappa}
\kappa_{i,i+1} = \frac{1}{2} \sqrt{(N-i)i},
\end{equation}
for $i=1,\cdots,N-1$. Using identical single-mode waveguides, it is possible to achieve this coupling by using different spacings between the waveguides. It is straightforward to show that the matrix defined in relation~(\ref{eq_F}) satisfies all properties of the discrete fractional Fourier transform, e.g., $F$ is unitary, $F^2$ is the parity operator, and $F^4$ is the identity matrix. Furthermore, for arbitrarily large $N$, the DFrFT defined in relation~(\ref{eq_F}) becomes a DFT. More importantly, this form of the DFrFT operation has a relatively simple physical realization as a waveguide array with nonuniform nearest neighbor coupling coefficients according to relation~(\ref{eq_kappa}) \cite{weimann2016implementation, HonariLatifpour2022}.

\begin{figure}[h!]
\centering
\includegraphics[width=0.5\textwidth]{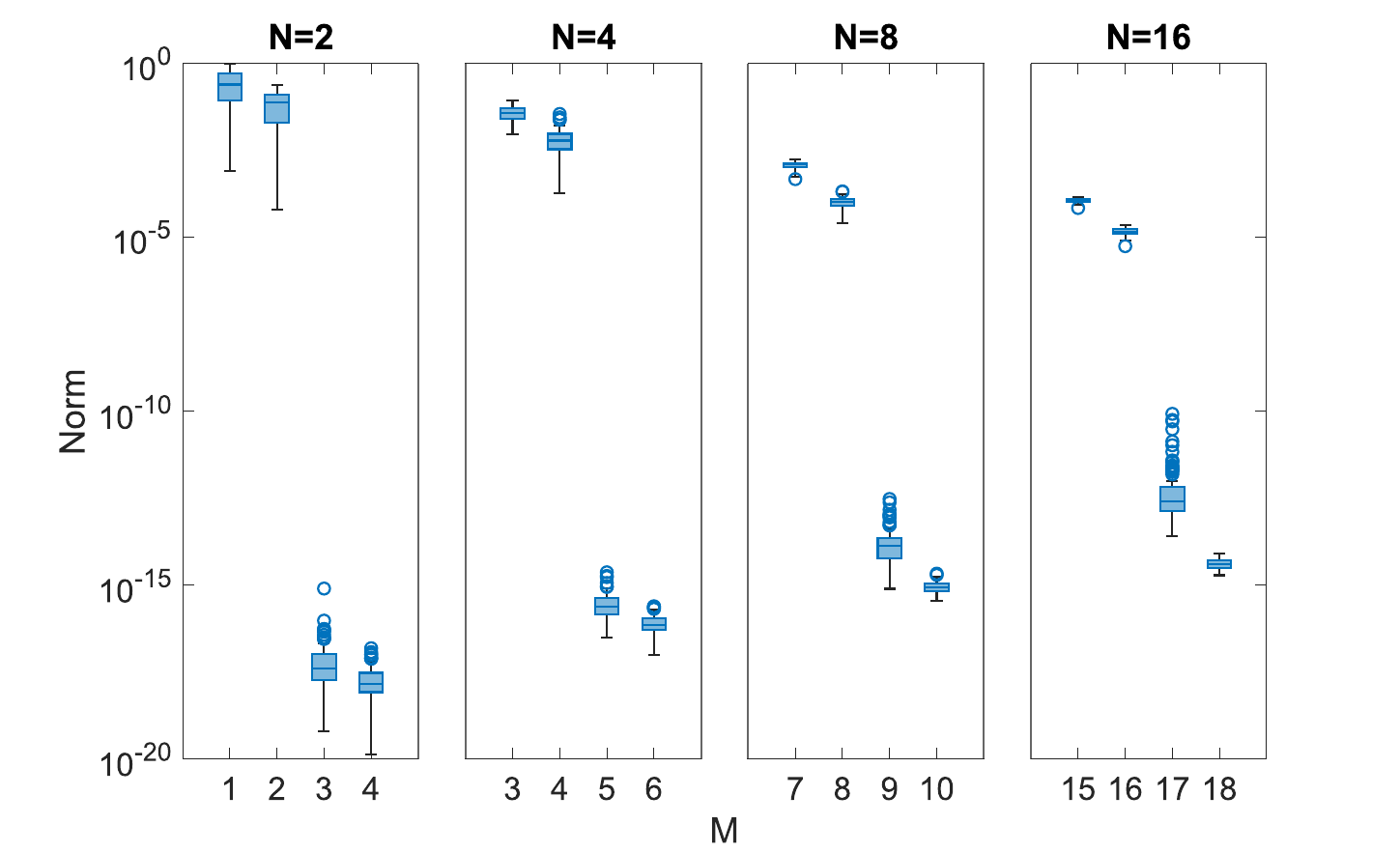}
\caption{The error norm (mean square error of the optimization) for randomly generated target unitary matrices of sizes $N=2, 4, 8,$ and $16$. In each case, $100$ target unitary matrices were generated, while for each target matrix, the optimization was run $100$ times with different initial conditions for different number of phase layers $M$ to get a single value of the lowest norm. In all cases, a phase transition is observed between $M=N$ and $M=N+1$.}
\label{fig2}
\end{figure}

To demonstrate the universality of this device, we optimize the individual phases for an ensemble of randomly chosen target unitary transformations $U_t$, generated in accordance with the Haar measure \cite{mezzadri_how_2007}. The goodness of approximation of the target matrices is explored against the number of layers $N$, with $M$ phase layers corresponding to $NM$ phase parameters. The loss function is defined as the mean square error
\begin{equation}
\label{eq_L}
L = \frac{1}{N^2} \| U - U_t \|^2,
\end{equation}
where $\|A\|=\sqrt{\text{Tr}(A^{\dagger}A)}$ is the Frobenius norm. Here we refer to Eq.~(\ref{eq_L}) as the error norm. The optimization is done using the Levenberg-Marquardt algorithm (LMA), which is well suited to sum-of-squares objective functions and which can be used for both under and over determined problems \cite{levenberg1944method}. The function tolerance and the step tolerance was set to $10^{-6}$ and the optimality tolerance to $10^{-10}$. 100 target unitary matrices were generated at each matrix size of $N=2,4,8,$ and $16$. For a given target, the phases were randomly initialized between $0$ and $2\pi$ and the LMA was run 100 times to find the parameters corresponding to the lowest error norm. For ($N=16, M=17$), it was found necessary to increase the number of runs to $1000$ to obtain the lowest possible error norms.

\begin{figure}[tb]
\includegraphics[width=0.42\textwidth]{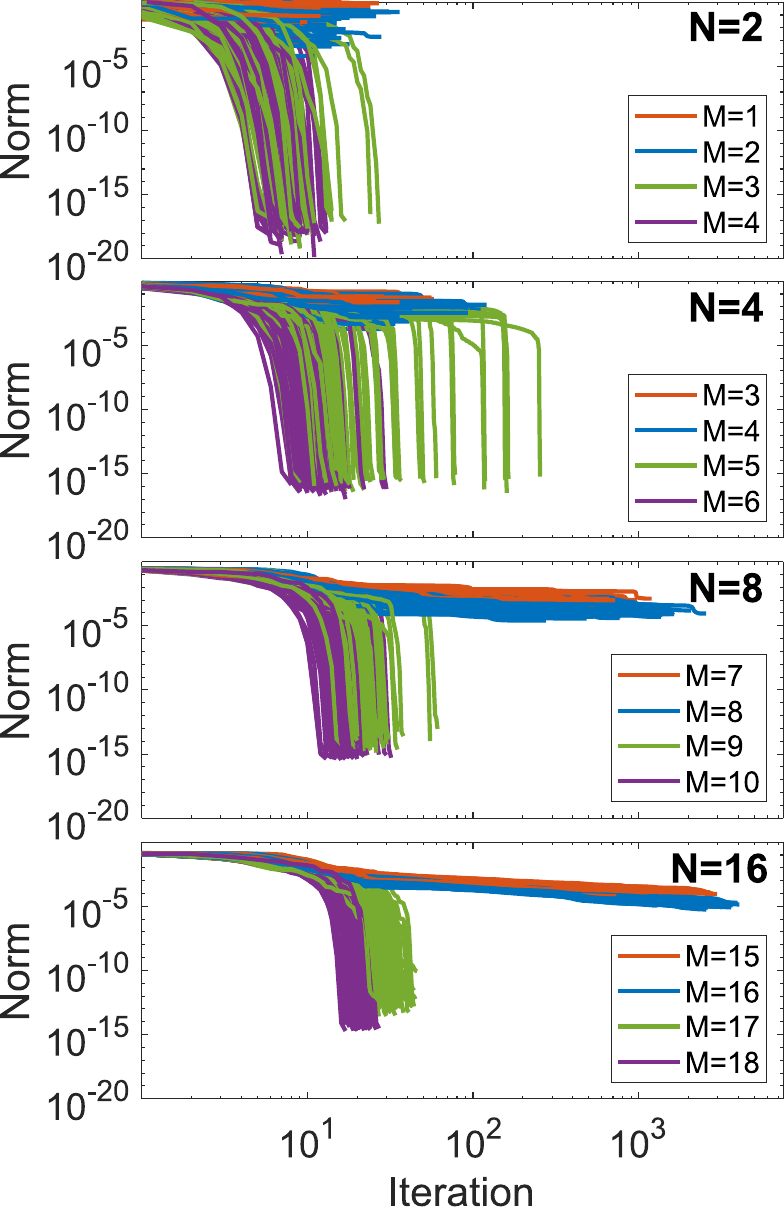}
\caption{The convergence of the error norm versus iteration step when considering different number of layers $M$, below and above the critical value, in the factorization. Here, for $N=2,4,8,$ and $16$, four values of $M$, $M=N-1, N, N+1,$ and $N+2$, are considered. In all graphs, only the run corresponding to the lowest norm for each target matrix and choice of $M$ is displayed.}
\label{fig3}
\end{figure}
\noindent

For a fixed relationship of the number of phase layers to the matrix order, it is seen in Fig.~\ref{fig2} that the error norm decreases as the latter increases. Depending on the desired accuracy, the special case $M=N$ may therefore be feasible for implementing universal circuits at larger N. Notably however, between $M=N$ and $M=N+1$ layers the error norm undergoes a phase transition. This suggests that $N+1$, rather than $N$ layers, corresponds to the more fundamental universality requirement, despite being slightly over-determined (with $N(N+1)$ parameters). Preliminary tests with a larger number of trials were performed to check that these gaps do not significantly close. With 10,000 trial runs for ($N = 2, M = 2$), the lowest error norm was found to be $10^{-7}$. These phase transitions can also be found using other popular methods such as Simulated Annealing and Genetic Algorithm. As these are probabilistic methods, a secondary optimization will be necessary to fine-tune the parameters in the ambient space so that the abrupt transition in the error norm can become visible.

To better investigate the phase transitions, the error norm versus the optimization iteration step is shown in Figs.~\ref{fig3}, once again using the best-case runs. For $M>N$, the norm of these runs varies slowly with the iteration step until a minimum is approached and there is an abrupt transition to a low value. In all cases, the error norm decreases as $M$ increases, with $M=N+2$ layers consistently performing well and with few iteration steps, corresponding to quick optimization times.

\begin{figure*}[h]
\centering
\includegraphics[width=0.7\textwidth]{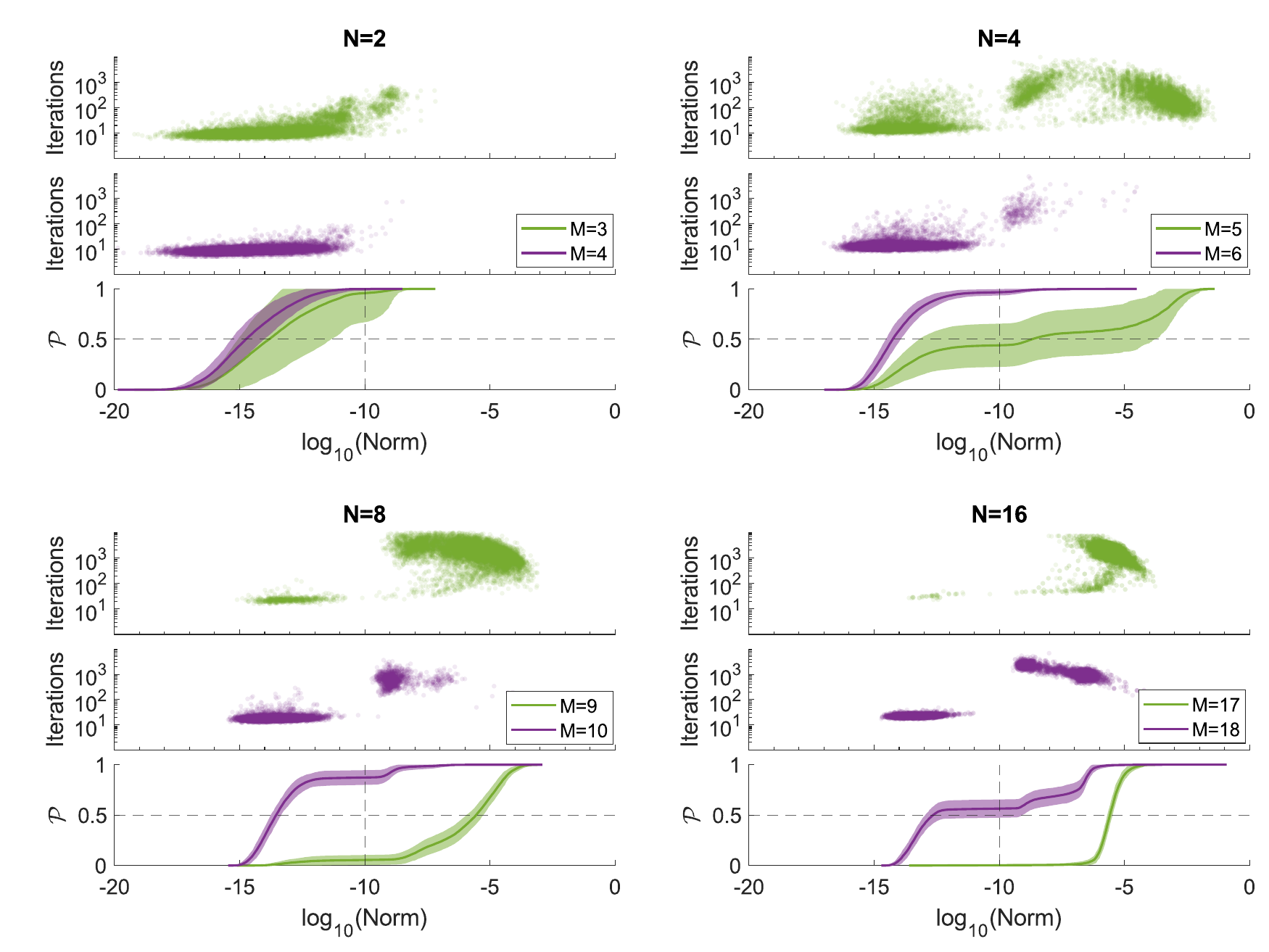}
\caption{(Top plots) The number of optimization iterations versus the norm of error for the $100 \times 100$ runs with $M=N+1$ and $M=N+2$, which can be used to estimate computation times. (Bottom plot) The cumulative probability of finding an error norm equal to or lower than a given value for a single run of LMA. Given that this cumulative probability depends on the target matrix, its distribution for many randomly generated target matrices is represented with shaded regions corresponding to $\pm 2$ standard deviations.}
\label{fig4}
\end{figure*}
\noindent

Clearly, to efficiently utilize the proposed architecture, it is important to have a reliable optimization scheme for finding the phase parameters that allow representing a desired unitary matrix. This is especially important for LMA, which can get stuck for long periods of time tuning the norm in small neighborhoods. Figure~\ref{fig4} shows the number of iterations and norms of the $100 \times 100$ runs for $M=N+1$ and $M=N+2$. As in Fig.~\ref{fig2}, there is clear evidence of a phase transition around $L_c = 10^{-10}$. Importantly, the number of iterations can differ by an order of magnitude or more depending on whether the norm of a single run converges below $L_c$. This suggests a modification of Levenberg-Marquardt with each run ending prematurely after a maximum iteration (or equivalently, time) threshold. The algorithm keeps performing trial runs until an error norm less than the critical value $L_c$ is achieved. The mean number of runs for this procedure can be deduced from the cumulative probability of the norms for a single run of LMA, shown in the bottom plots of Fig.~\ref{fig4}. The probability for a given norm decreases with $N$, but at $N=16$, there is still a $~50\%$ chance that with $M=N+2$ phase layers a norm of $L_c$ or lower will be achieved after a single run. For $M=N+1$, this probability is already low at $N=8$ ($\sim 1 \%$), leading to much greater computation times. Extrapolating from these trends for $N>16$, $M=N+3$ or higher may be required to achieve similar results. Thus, although the choice of $M=N+1$ phase layers has been shown to allow universality, there is a limit on the practicality of finding the solutions for very large N, even with truncated LMA. However, we stress again that it may not be required in practice to find these sub-$L_c$ solutions. If for example an error norm of $10^{-5}$ is considered good enough, then at $N=16$ there is a $\sim 90\%$ chance of obtaining a lower norm using untruncated LMA.

In summary, we proposed a novel architecture for the implementation of discrete linear unitary operators with photonic integrated circuits. The proposed architecture is built on interlacing fixed discrete fractional Fourier transform layers with programmable phase shift layers. Our results indicate that this architecture can universally represent $N \times N$ unitary operations with at most $N+1$ phase layers. The proposed architecture offers a fairly simple physical realization by utilizing photonic waveguide arrays in conjunction with phase shifter arrays.\\

\noindent
\textbf{Funding}. This project is supported by the U.S. Air Force Office of Scientific Research (AFOSR) Young Investigator Program (YIP) Award
FA9550-22-1-0189.194 \\

\noindent
\textbf{Disclosures}. Patent Pending

\bibliographystyle{pisikabst}
\bibliography{bibfile}

\end{document}